\documentclass[onecolumn,amsmath,amssymb,pra]{revtex4-2}

\usepackage{graphicx}
\usepackage{color}
\usepackage{dcolumn}
\usepackage{amsmath}
\usepackage{amsthm}
\usepackage{CJK}
\usepackage{subfigure}
\usepackage{sidecap}

\usepackage{amssymb}
\usepackage{mathrsfs}
\usepackage{amsfonts}
\usepackage{indentfirst}
\usepackage{bm,bbm}

\newtheorem{definition}{Definition}
\newtheorem{lemma}{Lemma}

\newtheorem{theorem}{Theorem}
\newtheorem{corollary}{Corollary}

\begin{document}

%
%



\title{Locality and entanglement of indistinguishable particles}
\author{Till Jonas Frederick Johann$^{1,2}$, Ugo Marzolino$^{2,3}$}
\affiliation{$^{1}$Ruprecht-Karls-Universit\"at Heidelberg, Germany}
\affiliation{$^{2}$Istituto Nazionale di Fisica Nucleare, Sezione di Trieste, Italy}

\begin{abstract}
Entanglement is one of the strongest quantum correlation, and is a key ingredient in fundamental aspects of quantum mechanics and a resource for quantum technologies. While entanglement theory is well settled for distinguishable particles, there are five inequivalent approaches to entanglement of indistinguishable particles. We analyse the different definitions of indistinguishable particle entanglement in the light of the locality notion. This notion is specified by two steps: \emph{i)} the identification of subsystems by means of their local operators; \emph{ii)} the requirement that entanglement represent correlations between the above subsets of operators. We prove that three of the aforementioned five entanglement definitions are incompatible with any locality notion defined as above.
\end{abstract}

\maketitle

\section{Introduction} \label{intro}

Entanglement is one of the strongest form of quantum correlations and is crucial for the undestanding of Bell's inequalities~\cite{BellGao}, quantum communications~\cite{Buhrman2010}, quantum metrology~\cite{Braun2018,Moreau2019}, and quantum computation~\cite{Lund2017,Harrow2017}.
Fundamental aspects of quantum theory and the quantum information framework suggest that entanglement is a notion derived from the definition of subsystems that can be correlated or, in other words, from the notion of \emph{locality}. For instance, any entangled pure state of two distinguishable particles violates a Bell's inequality, and thus prove quantum non-locality~\cite{Gisin1991,Gisin1992}, and the resourcefulness of Bell non-locality is a special instance of entanglement theory~\cite{Sengupta2020}.
On the other hand, several quantum technologies consist of parties independently manipulating subsystems, and entanglement among subsystems is a fundamental resource that allows to overcome the limitations of local operations on subsystems and classical communications between them~\cite{Chitambar2014}.

Entanglement theory is very well understood for distinguishable particles: two-particle separable, namely non-entangled, pure states are of the form $|\psi_1\rangle\otimes|\psi_2\rangle$, and each subsystem is implicitly assumed to be a particle. This definition can be reformulated by describing each subsystem with operators, termed \emph{local} and acting non-trivially only on it, i.e. $A=O_1\otimes\mathbbm{1}_2$ for the first particle and $B=\mathbbm{1}_1\otimes O_2$ for the second one. Therefore, a pure state is separable if and only if its expectations do not show correlations between single-particle operators: $\langle AB\rangle=\langle A\rangle\langle B\rangle$ for any $A,B$ as above~\cite{Werner1989}.

The choice of the operators $A$ and $B$ naturally arises from the experimental ability to individually address particles. Nevertheless, correlations between operators of different form have been proven to correspond to entanglement between more general subsystems~\cite{Zanardi2001,Zanardi2004,Barnum2004,Harshman2011,Thirring2011}, already in the framework of distinguishable particles.

\begin{figure*} [htbp]
\centering
\setlength{\fboxrule}{1.5pt}
\setlength{\fboxsep}{0pt}
\includegraphics[width=0.8\textwidth]{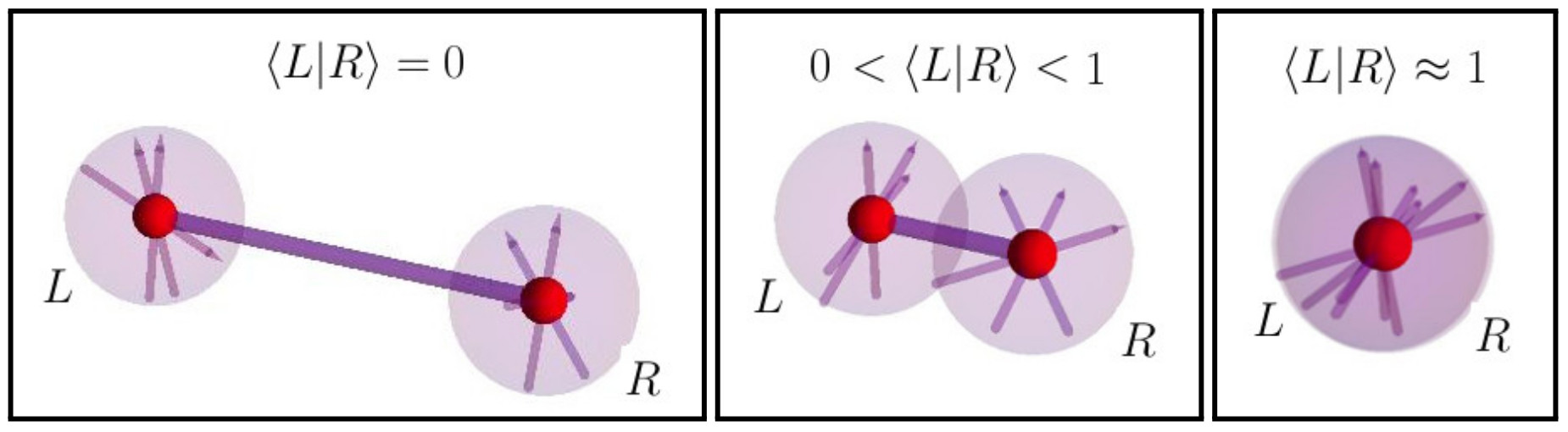}
\caption{Pictorial representation of two identical particles with an internal degree of freedom and a spatial degree of freedom, when the spatial wave function do not overlap $\langle L|R\rangle=0$ (effectively distinguishable particles), partially overlap $0<\langle L|R\rangle<1$, and overlap almost completely $\langle L|R\rangle\approx1$.}
\label{fig}
\end{figure*}

Although distinguishable particles are a very useful paradigm in many physical systems, e.g. spin models where particles are localised at different positions, Nature is made of several kinds of indistinguishable particles, like electrons, atoms and photons. Particle indistinguishability requires that pure states and operators be invariant under particle permutations. In particular, the aforementioned single-particle operators are no longer allowed. These considerations challenge the notion of particle as a natural subsystem, whenever indistinguishability cannot be neglected, e.g. if particles are not spatially separated (see figure \ref{fig} for an illustration of particles that are progressively less separated and lose their distinguishability).

In this context, five alternative definitions of entanglement  in systems of indistinguishable particles have been proposed: see~\cite{review} for a review. Nevertheless, the identification of subsystems and of local operators, whose correlations correspond to entanglement, is not always addressed.
This identification indeed allows us
to systematically describe the advantage that indistinguishable particle entanglement provides over local operations and classical communications in quantum technologies.
Other resource theories can
characterize resourceful states that overcome the limitations of operations defined by lifting the locality requirement.
The characterisation of local operators also enables us
to understand the overlap of a resource theory with entanglement theory, and the cost of resource conversion in terms of resourceful operations of both theories.

We say that a definition of entanglement is compatible with a notion of locality if entanglement corresponds to correlations between commuting subsets of operators $\mathcal{A}$ and $\mathcal{B}$: $\langle AB\rangle=\langle A\rangle\langle B\rangle$ for all $A\in\mathcal{A}$ and $B\in\mathcal{B}$
for pure separable states (see~\cite{Werner1989,review} for the generalization to mixed states). These operator subsets formally and operationally define subsystems as they consist of observables individually measurable and transformations induvidually implementable without mutual disturbance due to their commutativity $[\mathcal{A},\mathcal {B}]=0$.
A preliminary comparative analysis among the existing entanglement approaches shows that only one of them is fully consistent~\cite{review}. In particular, three of these approaches cannot be interpreted as entanglement of particles,
because they are not consistent with the correlations between
subsets made of
permutationally invariant single-particle operators.
The physical intuition is that truly indistinguishable particles cannot be individually addressed (see the sketch in Fig. \ref{fig}). This issue does not apply to the other approaches, since one,
the so-called superselection rule (SSR) entanglement, is resticted to physical situations of particles that
can be effectively distinguished by means of certain degrees of freedom~\cite{Wiseman2003,Ichikawa2010,Sasaki2011}, and the other accounts for more general correlations between modes in a second quantised description that also recovers particle correlations for
effectively distinguishable particles~\cite{Zanardi2002-1,Shi2003,Schuch2004-2,Benatti2010,Benatti2012,Benatti2012-2,
Marzolino2013,Benatti2014,Balachandran2013-2,Benatti2016}.
The latter notion, i.e. mode-entanglement, is also routinely applied for entanglement detection, manipulations and measures~\cite{Cramer2013,Killoran2014,Ding2020,review}.

In this paper, we focus on the aforementioned entanglement definitions that do not correspond to particle correlations, and investigate if they rather represent correlations between subsystems identified by general subsets $\mathcal{A}$ and $\mathcal{B}$.
For each definition, we define the set of separable pure states, namely $\mathsf{SEP}$, and look for candidates for the subsets $\mathcal{A}$ and $\mathcal{B}$.

Before sketching the general scheme of our analysis, it is crucial to notice that the linearity of quantum mechanics implies that the set of operators of a system, as well as subsystems, is an algebra~\cite{BratteliRobinson,Strocchi}.
An algebra is a linear space closed under a multiplication between its elements and under a conjugation operation (hermitian conjugation in our case).
Therefore, the subsets $\mathcal{A}$ and $\mathcal{B}$ that comprise all operators acting on each subsystem are algebras, as happens in the standard case of distinguishable particles.
Nevertheless, we have used neither the algebra structure nor commutativity in our main results which therefore hold also for more general subsets of operators, as those considered in~\cite{Kabernik2020}.

A necessary condition for operators $A\in\mathcal{A}$ and $B\in\mathcal{B}$ is that they do not generate entanglement, because subsystems cannot be correlated by local operators, like $A$, $B$, and $AB$.
This requirement is formulated in full generality as $A\cdot\mathsf{SEP}=\mathsf{SEP}$ and $B\cdot\mathsf{SEP}=\mathsf{SEP}$, or
\begin{equation} \label{ent-gen}
A\,|\Psi\rangle,\,\,\,B\,|\Psi\rangle\in\mathsf{SEP}, \qquad \forall\,\,|\Psi\rangle\in\mathsf{SEP}
\end{equation}
relaxing the normalisation conditions, such that $\big|\big||\Psi\rangle\big|\big|>0$, $\big|\big|A|\Psi\rangle\big|\big|>0$, $\big|\big|B|\Psi\rangle\big|\big|>0$.

Operators in each subset, $\mathcal{A}$ and $\mathcal{B}$, are chosen from the above ones, and those belonging to different subsets must commute with each other. Thus, we check the factorisation condition
\begin{equation} \label{fact}
\frac{\langle\Psi|AB|\Psi\rangle}{\langle\Psi|\Psi\rangle}=\frac{\langle\Psi|A|\Psi\rangle}{\langle\Psi|\Psi\rangle}\frac{\langle\Psi|B|\Psi\rangle}{\langle\Psi|\Psi\rangle}, \qquad
\forall \quad A\in\mathcal{A}, \quad B\in\mathcal{B}, \quad |\Psi\rangle\in\mathsf{SEP}
\end{equation}
for any subset, $\mathcal{A}$ and $\mathcal{B}$, of operators that do not generate entanglement.

\section{Results} \label{results}

We are now ready to go through the different entanglement definitions and to look for possible subsets, $\mathcal{A}$ and $\mathcal{B}$, of local operators.
We shall prove that it is impossible to find subsets whose correlations correspond to the following notions of entanglement.
Specifically, we find that for any possible subset bipartition there are separable states that do not fulfil the factorisation of local expectations \eqref{fact}.

The first quantisation formalism is most familiar to the entanglement definitions analysed here. For completeness, we provide some definitions both in first and second quantisation, using the tilde to distinguish the second quantization formalism. However, our computations rely on matrix representations of the Hilbert space that are equivalent to matrix representations of Fock space sectors, after rewriting the basis states from first to second quantisation. Therefore, our conclusions remain valid in both formalisms, and we often consider the symbols without the tilde for simplifying the notation.

Moreover, we shall present our results in the form of lemmas and theorems in order to emphasise the key results in the statements.

\subsection{Entanglement-I} \label{entI}

For the sake of concreteness, let us focus on two bosonic two-level particles. 
Define a basis of the single particle Hilbert space $\mathbbm{C}^2$, namely $|0\rangle$ and $|1\rangle$, and that of the symmetrised two-particle Hilbert space $\mathsf{S}\big(\mathbbm{C}^2\otimes\mathbbm{C}^2\big)$ where $\mathsf{S}$ is the symmetrisation projector,
\begin{equation} \label{basis}
|\Phi_0 \rangle=|0\rangle|0\rangle,
\qquad
|\Phi_1 \rangle=|1\rangle|1\rangle,
\qquad
|\Phi_2 \rangle=\frac{|0\rangle|1\rangle + |1\rangle|0\rangle}{\sqrt{2}}.
\end{equation}
In second quantisation the above states read

\begin{equation} \label{basis2}
|\widetilde\Phi_0 \rangle=\frac{\big(\mathfrak{a}_0^\dag\big)^2}{\sqrt{2}}|\textnormal{vac}\rangle,
\qquad
|\widetilde\Phi_1 \rangle=\frac{\big(\mathfrak{a}_1^\dag\big)^2}{\sqrt{2}}|\textnormal{vac}\rangle,
\qquad
|\widetilde\Phi_2 \rangle=\mathfrak{a}_0^\dag\mathfrak{a}_1^\dag|\textnormal{vac}\rangle,
\end{equation}
where $\mathfrak{a}_{0,1}^\dag$ are creation operators of a single particle in the state $|0\rangle$ or $|1\rangle$ respectively, with $[\mathfrak{a}_i,\mathfrak{a}_j^\dag]=\delta_{i,j}$, and $|\textnormal{vac}\rangle$ is the vacuum.

In this case, the first entanglement definition is

\begin{definition}[Entanglement-I~\cite{Paskauskas2001,Eckert2002,Grabowski2011,Morris2020}] \label{defI}
The set of
pure separable-I states
is, in first and second quantization respectively,
\begin{align}
\mathsf{SEP}_{\mathsf{I}} & =\Big\{\big(c_0 |0\rangle+c_1 |1\rangle\big)^{\otimes 2}\Big\}_{c_0,c_1\in\mathbbm{C}}=\big\{c_0^{2}|\Phi_0\rangle+c_1^{2}|\Phi_1\rangle+\sqrt{2}\,c_0 c_1 |\Phi_2 \rangle\big\}_{c_0,c_1\in\mathbbm{C}},
\label{SI} \\
\widetilde{\mathsf{SEP}}_{\mathsf{I}} &
=\Big\{\frac{1}{\sqrt{2}}\,\big(c_0 \, \mathfrak{a}_0^\dag+c_1 \, \mathfrak{a}_1^\dag\big)^2|\textnormal{vac}\rangle\Big\}_{c_0,c_1\in\mathbbm{C}}
=\big\{c_0^{2}|\widetilde\Phi_0\rangle+c_1^{2}|\widetilde\Phi_1\rangle+\sqrt{2}\,c_0 c_1 |\widetilde\Phi_2 \rangle\big\}_{c_0,c_1\in\mathbbm{C}}.
\label{SI2}
\end{align}
All other pure states are entangled-I.
\end{definition}

Practical realisations of the above single particle states $|0\rangle$ and $|1\rangle$ are, e.g. in cold atoms, spatial localisation in separated wells of a lattice, or hyperfine energy levels. Therefore, separable-I states are those with particles in the same superposition of localised or energy states.

In the following theorem, we characterise operators that do not generate entanglement-I.

\begin{theorem}  \label{theoremI1}
Operators that leave $\mathsf{SEP}_{\mathsf{I}}$ invariant are
represented in first quantisation as $O\otimes O$ on the enlarged Hilbert space $\mathbbm{C}^2\otimes\mathbbm{C}^2$ with $O=O^\dag$, and in second quantisation as
\begin{equation} \label{loc.opI-IIQ}
\frac{1}{2}\sum_{i,j,k,l=0,1}\langle i|O|j\rangle\langle k|O|l\rangle \, \mathfrak{a}^\dag_i\mathfrak{a}^\dag_k\mathfrak{a}_j\mathfrak{a}_l.
\end{equation}
\end{theorem}
\begin{proof}
Any operator $A$ that does not generate entanglement-I fulfils equation \eqref{ent-gen} which implies, from Definition \ref{defI},

\begin{equation} \label{square}
\langle\Phi_2|A|\Psi\rangle^2=2 \, \langle\Phi_0|A|\Psi\rangle\langle\Phi_1|A|\Psi\rangle.
\end{equation}
where $|\Psi\rangle\in\mathsf{SEP}_{\mathsf{I}}$.
Note that only the projection $\mathsf{S}A\mathsf{S}$ onto the symmetric space $\mathsf{S}\big(\mathbbm{C}^2\otimes\mathbbm{C}^2\big)$ contribute to equation \eqref{AI} since we considered bosonic states.
Equalities \eqref{square} for $|\Psi\rangle=|\Phi_{0,1}\rangle$ imply
\begin{equation} \label{AI}
\mathsf{S}A\mathsf{S}=
\begin{pmatrix}
a_{00}^2 & a_{10}^2 & x \\
a_{01}^{2} & a_{11}^{2} & y \\
\sqrt{2}\,a_{00}\,a_ {01} & \sqrt{2}\,a_{10}\,a_{11} & z
\end{pmatrix},
\end{equation}
in the basis \eqref{basis}, or \eqref{basis2}, with $a_{00,01,10,11},x,y,z\in\mathbbm{C}$.

Conditions for $x$ and $y$ are found by requiring equation \eqref{ent-gen} for arbitrary $|\Psi\rangle\in\mathsf{SEP}_{\mathsf{I}}$. This requirement implies, after plugging into \eqref{square} the explicit form of $|\Psi\rangle$ in equation \eqref{SI}, that the coefficients of each monomial $c_0^nc_1^m$ vanish. We obtain
\begin{align}
&z^2+2\,a_{00}\,a_{01}\,a_{10}\,a_{11}=a_{00}^2\,a_{11}^2+a_{01}^2\,a_{10}^2+2xy,
\label{eq:ent_1_matrix_eq_square}\\
&\sqrt{2}\,a_{00}\,a_{01}\,z=a_{00}^2\,y+a_{01}^2\,x,
\label{eq:ent_1_matrix_eq_1}\\
&\sqrt{2}\,a_{10}\,a_{11}\,z=a_{10}^2\,y+a_{11}^2\,x.
\label{eq:ent_1_matrix_eq_2}
\end{align}
The solutions of the above equations are
\begin{align}
x= & \pm\sqrt{2}\,a_{00}\,a_ {10}, \\
y= & \pm\sqrt{2}\,a_{01}\,a_ {11}, \\
z= & \pm a_{00}\,a_ {11}\pm a_{01}\,a_ {10},
\end{align}
with either all plus signs or all minus signs.

The two cases, e.g. plus and minus signs, are equivalent as they result in the same operator by redefining $a_{00}'=-a_{00}$ and $a_{01}'=-a_{01}$. Therefore, we found
\begin{equation} \label{AlocI}
\mathsf{S}A\mathsf{S}=
\begin{pmatrix}
a_{00}^2 & a_{10}^2 & \sqrt{2}\,a_{00}\,a_ {10} \\
a_{01}^{2} & a_{11}^{2} & \sqrt{2}\,a_{01}\,a_{11} \\
\sqrt{2}\,a_{00}\,a_ {01} & \sqrt{2}\,a_{10}\,a_{11} & a_{00}\,a_{11}+a_{01}\,a_{10}
\end{pmatrix}.
\end{equation}

In first quantisation, the above matrix is exactly the projection onto the symmetric subspace of the operator $O\otimes O$ on the larger Hilbert space $\mathbbm{C}^2\otimes\mathbbm{C}^2$, with
\begin{equation}
O=
\begin{pmatrix}
a_{00} & a_{10} \\
a_{01} & a_{11}
\end{pmatrix},
\end{equation}
in the basis $\big\{|0\rangle,|1\rangle\big\}$.
In second quantization, equation \eqref{AlocI} is the matrix representation of equation \eqref{loc.opI-IIQ}.
This concludes the proof.
\end{proof}

The factorisation condition can be checked for arbitrary operators, $A=O\otimes O$ and $B=Q\otimes Q$, that could form subsets of local operators. Note that the subsets are not necessarily commuting in the next theorem. We have therefore proven a more general result than what we need in entanglement analysis.

\begin{theorem} \label{theoremI3}
Given two subsets $\mathcal{A}$ and $\mathcal{B}$ of operators that leave $\mathsf{S}_{\textnormal{sep}}^I$ invariant, the factorisation condition \eqref{fact}
and Definition \ref{defI} imply that either $\mathcal{A}$ or $\mathcal{B}$ consists only of operators proportional to the identity.
\end{theorem}
\begin{proof}
Consider a separable-I state $|\Psi\rangle\in\mathsf{SEP}_{\mathsf{I}}$ as in Definition \ref{defI}, and two operators, $A\in\mathcal{A}$ and $B\in\mathcal{B}$ as in Theorem \ref{theoremI1}. The factorisation condition \eqref{fact} reads


\begin{equation} \label{fact.entI2}
\left(\sum_{j=0,1}|c_j|^2\right)^2\left(\sum_{l,k=0,1}\langle l|OQ|k\rangle\,\overline{c_l}\,c_k\right)^2
=\left(\sum_{i,j,l,k=0,1}\langle i|O|j\rangle\langle l|Q|k\rangle\,\overline{c_i}\,c_j\,\overline{c_l}\,c_k\right)^2.
\end{equation}
Both the left and the right hand sides of equation \eqref{fact.entI2} are eight order polynomials in $c_0$ and $c_1$. Nevertheless, there are monomials in the right-hand-side that do not appear in the left-hand-side. Since the factorisation condition must hold for any separable-I state, thus for any $c_0$ and $c_1$, the monomials that appear only in the right-hand-side, i.e. with $i\neq j$ and $l\neq k$, must by multiplied by vanishing coefficients, i.e. $\langle i|O|j\rangle\langle l|Q|k\rangle=0$. This consideration implies that either $O$ or $Q$ are diagonal in the basis $\{|0\rangle,|1\rangle\}$. The latter basis is arbitrary and the diagonality requirement must hold for any basis rotation. In conclusion, either $O$ or $Q$ must be proportional to the identity.
\end{proof}

Therefore, entanglement-I is incompatible with any locality notions where each subsystem is characterised by subsets of local operators. We have not used the commutativity of subsets $\mathcal{A}$ and $\mathcal{B}$ in Theorem \ref{theoremI3}. In the Supplementary Information, we provide a characterisation of commuting subsets of operators that do not generate entanglement-I, a second proof of Theorem \ref{theoremI3} using $[\mathcal{A},\mathcal{B}]=0$, and some examples of the proven properties.

\subsection{Entanglement-II} \label{entII}

Considering again two bosonic two-level particles, the second entanglement definition is

\begin{definition}[Entanglement-II~\cite{Herbut2001,Schliemann2001,Paskauskas2001,Eckert2002,
Plastino2009,
Grabowski2011,Li2001,
Ghirardi2002,
Iemini2013}] \label{defII}
The set of
pure separable-II states
is, in first and second quantization respectively,
\begin{align}
\mathsf{SEP}_{\mathsf{II}} & =
\mathsf{SEP}_{\mathsf{I}}\cup
\Big\{\mathsf{S}\big(|c_0 |0\rangle+c_1 |1\rangle\big)\otimes\big(|\overline{c_1}|0\rangle-\overline{c_0}|1\rangle\big)\Big\}_{c_{0,1}\in\mathbbm{C}}
=\mathsf{SEP}_{\mathsf{I}}\cup
\left\{c_0\overline{c_1}|\Phi_0\rangle-c_1\overline{c_0}|\Phi_1\rangle+\frac{|c_1 |^2-|c_0|^2}{\sqrt{2}}|\Phi_2 \rangle\right\}_{c_{0,,1}\in\mathbbm{C}} \\
\widetilde{\mathsf{SEP}}_{\mathsf{II}} &
=\widetilde{\mathsf{SEP}}_{\mathsf{I}}\cup
\Big\{\big(c_0 \, \mathfrak{a}^\dag_0+c_1 \, \mathfrak{a}^\dag_1\big)\big(\bar c_1 \, \mathfrak{a}^\dag_0-\bar c_0 \, \mathfrak{a}^\dag_1\big)|\textnormal{vac}\rangle\Big\}_{c_{0,1}\in\mathbbm{C}}
=\widetilde{\mathsf{SEP}}_{\mathsf{I}}\cup
\left\{c_0\overline{c_1}|\widetilde\Phi_0\rangle-c_1\overline{c_0}|\widetilde\Phi_1\rangle+\frac{|c_1 |^2-|c_0|^2}{\sqrt{2}}|\widetilde\Phi_2 \rangle\right\}_{c_{0,1}\in\mathbbm{C}}
\end{align}
All other pure states are entangled-II. 
\end{definition}

Keeping in mind the same physical realisations mentioned after Definition \ref{defI}, two particles are separable-II if they are either in the same or in orthogonal superpositions of spatial localisations or of hyperfine levels.

\begin{theorem} \label{theoremII1}
Any operator that leaves $\mathsf{SEP}_{\mathsf{II}}$ invariant also leaves $\mathsf{SEP}_{\mathsf{I}}$ invariant.
\end{theorem}
\begin{proof}
Consider an operator $A$ that does not generate entanglement-II, and therefore leaves $\mathsf{SEP}_{\mathsf{II}}$ invariant, and a state $|\Psi\rangle\in\mathsf{SEP}_{\mathsf{I}}\subset\mathsf{SEP}_{\mathsf{II}}$. Either $A|\Psi\rangle\in\mathsf{SEP}_{\mathsf{I}}$ or $A|\Psi\rangle\in\mathsf{SEP}_{\mathsf{II}}\setminus\mathsf{SEP}_{\mathsf{I}}$. A necessary and sufficient condition for $A|\Psi\rangle\in\mathsf{SEP}_{\mathsf{I}}$ is equation \eqref{square}, namely

\begin{equation} \label{PA}
P_A(c_0,c_1):=\langle\Phi_2|A|\Psi\rangle^2-2\,\langle\Phi_0|A|\Psi\rangle\langle\Phi_1|A|\Psi\rangle=0.
\end{equation}
Instead, a necessary and sufficient condition for $A|\Psi\rangle\in\mathsf{SEP}_{\mathsf{II}}\setminus\mathsf{SEP}_{\mathsf{I}}$, in the case of two bosonic two-level particles, is~\cite{Ghirardi2004}

\begin{equation} \label{QA}
Q_A(c_0,\overline{c_0},c_1,\overline{c_1})=\textnormal{Tr}_1\left(\textnormal{Tr}_2\frac{A|\Psi\rangle\langle\Psi|A^\dag}{\langle\Psi|A^\dag A|\Psi\rangle}\right)^2-\frac{1}{2}=0,
\end{equation}
where $\textnormal{Tr}_j$ is the standard partial trace over the $j$-th particle Hilbert space. This partial trace does not have a straightforward connection with entanglement when indistinguishability cannot be neglected due to the symmetrisation postulate, as discussed here and in~\cite{review}. Nevertheless, it is a mathematical operation that still provides information on the structure of states.

$P_A(c_0,c_1)$ is a forth order homogeneous polynomial in $c_0$ and $c_1$ but does not depend on their complex conjugates.
The fundamental theorem of algebra states that, for any fixed $c_0$, equation \eqref{PA} either has at least one up to four distinct solutions, $c_1=c_1(c_0)$, or is a tautology.

$Q_A(c_0,\overline{c_0},c_1,\overline{c_1})$ is an eighth order polynomial that depends also on the complex conjugates $\overline{c_{0,1}}$, so that the fundamental theorem of algebra does not apply. Equation \eqref{QA} can be a tautology, have at most eight solutions, say $\textnormal{Re}\,c_1(c_0,\textnormal{Im}\,c_1)$ for fixed $c_0$ and $\textnormal{Im}\,c_1$, but can also have no solutions. If equation \eqref{QA} were a tautology, all states in $\mathsf{SEP}_{\mathsf{I}}$ are transformed into states in $\mathsf{SEP}_{\mathsf{II}}\setminus\mathsf{SEP}_{\mathsf{I}}$. Nevertheless, equation \eqref{PA} has at least a solution, and so at least one state in $\mathsf{SEP}_{\mathsf{I}}$ is transformed into a state in $\mathsf{SEP}_{\mathsf{I}}$ but also in $\mathsf{SEP}_{\mathsf{II}}\setminus\mathsf{SEP}_{\mathsf{I}}$. This result is impossible because of the linearity of $A$, unless $A|\Psi\rangle=0$ for all $|\Psi\rangle$ that solve equation \eqref{PA}. The latter case implies $A=0$, because the solutions of equation \eqref{PA} span the full Hilbert space $\mathsf{S}\big(\mathbbm{C}^2\otimes\mathbbm{C}^2\big)$, as proven in the Lemma \ref{theorem0} below. Thus, equation \eqref{QA} has at most eight solutions $\textnormal{Re}\,c_1(c_0,\textnormal{Im}\,c_1)$ for fixed $c_0$ and $\textnormal{Im}\,c_1$.

The above argument shows that, fixing $c_0$ and $\textnormal{Im}\,c_1$, there are at most a finite number of states $|\Psi\rangle\in\mathsf{SEP}_{\mathsf{I}}$ (corresponding to coefficients $(c_0,c_1)$) that are sent to $\mathsf{SEP}_{\mathsf{II}}\setminus\mathsf{SEP}_{\mathsf{I}}$.
If equation \eqref{PA} has at most four solutions $c_1(c_0)$ for fixed $c_0$, there is also at most a finite number of states $|\Psi\rangle\in\mathsf{SEP}_{\mathsf{I}}$ that are sent to $\mathsf{SEP}_{\mathsf{I}}$.
Remind, however, that we have relaxed the normalisation condition $\langle\Psi|\Psi\rangle=1$, and so $\textnormal{Re}\,c_1$ can assume a continuity of values. Therefore, there are states $|\Psi\rangle\in\mathsf{SEP}_{\mathsf{I}}$ that are sent out of $\mathsf{SEP}_{\mathsf{II}}$, and this contradicts the hypothesis of Theorem \ref{theoremII1}.

In conclusion, the only possibility is that equation \eqref{PA} is a tautology and equation \eqref{QA} has no solutions. This is equivalent to the statement of Theorem \ref{theoremII1}.
\end{proof}

\begin{lemma} \label{theorem0}
The states $|\Psi\rangle$ that solve equation \eqref{PA} span the whole symmetrised Hilbert space $\mathsf{S}\big(\mathbbm{C}^2\otimes\mathbbm{C}^2\big)$.
\end{lemma}
\begin{proof}
If equation \eqref{PA} is a tautology, its solutions correspond to all states $|\Psi\rangle\in\mathsf{SEP}_{\mathsf{I}}$ which form an overcomplete basis of the symmetrised Hilbert space $\mathsf{S}\big(\mathbbm{C}^2\otimes\mathbbm{C}^2\big)$. Therefore the statement is proved.

If equation \eqref{PA} is not a tautology, it has at least one and at most four solutions $c_1=f(c_0)$ for each fixed $c_0$ and fixed $A$, from the fundamental theorem of algebra. Denote by $|\Psi(c_0)\rangle\in\mathsf{SEP}_{\mathsf{I}}$ the state corresponding to one of these solutions with $c_0\neq0$. Consider the rescaling $c_0'=\lambda c_0$, and assume that the new state $|\Psi(c_0')\rangle$ is proportional to $|\Psi(c_0)\rangle$. Therefore $|\Psi(c_0')\rangle=\lambda^2|\Psi(c_0)\rangle$ and $c_1'=\lambda c_1$.
We implicitly assumed that $c_1=f(c_0)$ and $c_1'=f(c_0')$ are the same function, which is always possible because the polynomial $P_A$ depends only on the matrix $A$ that is not changed.

Equations $f(\lambda c_0)=f(c_0')=c_1'=\lambda c_1=\lambda f(c_0)$ say that the function $c_1=f(c_0)$ is homogeneous of degree one and therefore linear. This linear behaviour implies that $P_A(c_0,c_1)\propto c_0^4$, because $P_A$ is homogeneous in $c_0$ and $c_1$. Therefore $P_A=0$ implies $c_0=0$, but $c_0\neq0$ by construction. Therefore, $|\Psi(c_0)\rangle$ and $|\Psi(c_0')\rangle$ are linearly independent. We can iterate this argument with a second rescaling $c_0''=\eta c_0'=\eta\lambda c_0$ to find a third state linearly independent from the previous two. This proves the Theorem, since the symmetrised Hilbert space $\mathsf{S}\big(\mathbbm{C}^2\otimes\mathbbm{C}^2\big)$ is three-dimensional.
\end{proof}

The impossibility to define subsets of local operators, whose correlations correspond to entanglement-II, follows from Theorem \ref{theoremII1} and from the analogous result for entanglement-I.

\begin{theorem} \label{theoremII3}
Given two subsets $\mathcal{A}$ and $\mathcal{B}$ of operators that leave $\mathsf{SEP}_{\mathsf{II}}$ invariant, the factorisation condition \eqref{fact}
and Definition \ref{defII} imply that either $\mathcal{A}$ or $\mathcal{B}$ consists only of operators proportional to the identity.
\end{theorem}
\begin{proof}
Theorem \ref{theoremII1} implies that operators that generate subsets $\mathcal{A}$ and $\mathcal{B}$ are special cases of those considered in Theorem \ref{theoremI1} and in Theorem \ref{theoremI3}. The statements in Theorem \ref{theoremI1} and in Theorem \ref{theoremI3} therefore holds also for the operators considered here. Moreover, separable-I states are also separable-II, so that Theorem \ref{theoremI3} proves the inconsistency also between separability-II and locality.
\end{proof}
Note that also the proof of Theorem \ref{theoremII3} does not require the commutativity of the operator subsets.
In the Supplementary Information, we prove a complete characterisation of operators that do not generate entanglement-II, and prove that their subsets do not form an algebra. This provides an alternative proof of Theorem \ref{theoremII3} if the operator subsets $\mathcal{A}$ and $\mathcal{B}$ are algebras. We also show some examples in the Supplementary Information.

\subsection{Entanglement-III} \label{entIII}

The last definition of entanglement is more elaborated than the previous ones.
The simplest physical system, where entanglement-III exists, is made
of two bosons with a spatial degree of freedom, $\textnormal{span}\{|L\rangle,|R\rangle\}$, and an internal one, $\textnormal{span}\{|0\rangle,|1\rangle\}$. This definition is relative to the projection onto a subspace of the single particle Hilbert space. We assume that such subspace is $\mathcal{K}=\textnormal{span}\big\{|L,\sigma\rangle\big\}_{\sigma=0,1}$, as often assumed within this approach~\cite{LoFranco2016}.
The general definition, shown in the Supplementary Information, leads in our case to the following.

\begin{definition}[Entanglement-III~\cite{LoFranco2016}]
\label{defIII}
The set of pure separable-III states is, in first and second quantization respectively,
\begin{align}
\mathsf{SEP}_{\mathsf{III}} & =\bigg\{a_{\sigma}\,|L,\sigma\rangle^{\otimes 2}+b_{\sigma,\sigma'}\,\mathsf{S}\,|L,\sigma\rangle\otimes|R,\sigma'\rangle
+\sum_{\sigma_1,\sigma_2=0,1}c_{\sigma_1,\sigma_2}\,\mathsf{S}\,|R,\sigma_1\rangle\otimes|R,\sigma_2\rangle\bigg\}_{\substack{|\sigma\rangle,|\sigma'\rangle\in\mathbbm{C}^2\\a_{\sigma},\,b_{\sigma,\sigma'},\,c_{\sigma_1,\sigma_2}\in\mathbbm{C}}} \\
\widetilde{\mathsf{SEP}}_{\mathsf{III}} &
=\bigg\{\bigg(a_{\sigma} \, \frac{\big(\mathfrak{a}_{L,\sigma}^\dag\big)^2}{\sqrt{2}}+b_{\sigma,\sigma'} \, \mathfrak{a}_{L,\sigma}^\dag\mathfrak{a}_{R,\sigma'}^\dag
+\sum_{\sigma_1,\sigma_2=0,1}c_{\sigma_1,\sigma_2} \, \mathfrak{a}_{R,\sigma_1}^\dag\mathfrak{a}_{R,\sigma_2}^\dag\bigg)|\textnormal{vac}\rangle\bigg\}_{\substack{|\sigma\rangle,|\sigma'\rangle\in\mathbbm{C}^2\\a_{\sigma},\,b_{\sigma,\sigma'},\,c_{\sigma_1,\sigma_2}\in\mathbbm{C}}}
\end{align}
All other pure states are entangled-III.
\end{definition}

Let us introduce the following projectors in first quantisation

\begin{equation} \label{proj}
P_X=\sum_{\sigma_1,\sigma_2=0,1}\big(2-\delta_{X_1,X_2}\delta_{\sigma_1,\sigma_2}\big)\,\mathsf{S}\big(|X_1,\sigma_1\rangle\langle X_1,\sigma_1|\otimes|X_2,\sigma_2\rangle\langle X_2,\sigma_2|\big)\mathsf{S},
\end{equation}
or in second quantisation

\begin{equation} \label{proj2}
\widetilde P_X=\frac{1}{2}\sum_{\sigma_1,\sigma_2=0,1}\mathfrak{a}_{X_1,\sigma_1}^\dag\mathfrak{a}_{X_2,\sigma_2}^\dag\mathfrak{a}_{X_2,\sigma_2}\mathfrak{a}_{X_1,\sigma_1}
\end{equation}
with $X=X_1X_2$, and $X_{1,2}\in\{L,R\}$, and define $A_{X,Y}=P_X\,A\,P_Y$ for any operator $A$. The support of $P_{LL}$ is isomorphic to $\mathsf{S}\big(\mathbbm{C}^2\otimes\mathbbm{C}^2\big)$, and separable-III states in $P_{LL}\cdot\mathsf{SEP}_{\mathsf{III}}$, namely $|L,\sigma\rangle^{\otimes 2}$, are in one-to-one correspondence with separable-I states $\mathsf{SEP}_{\mathsf{I}}$
in equation \eqref{SI}. In particular, all the results for entanglement-I hold for entanglement-III constrained to the support of $P_{LL}$. The support of $P_{LR}$ is isomorphic to $\mathbbm{C}^2\otimes\mathbbm{C}^2$~\cite{DeMuynck1975,Herbut2001,Herbut2006,Tichy2013,Cunden2014,
review}; therefore, separable-III states in $P_{LR}\cdot\mathsf{SEP}_{\mathsf{III}}$, i.e. $\mathsf{S}\,|L,\sigma\rangle\otimes|R,\sigma'\rangle$, are in one-to-one correspondence with two distinguishable two-level particles, $|\sigma\rangle\otimes|\sigma'\rangle=\big(c_0|0\rangle+c_1|1\rangle\big)\otimes\big(d_0|0\rangle+d_1|1\rangle\big)$. Finally, $P_{RR}\cdot\mathsf{SEP}_{\mathsf{III}}$ is the whole support of $P_{RR}$ and isomorphic to $\mathsf{S}\big(\mathbbm{C}^2\otimes\mathbbm{C}^2\big)$.

The search for subsets of local operators requires a detailed analysis of contributions to the factorisation condition from each sector identified by the projectors \eqref{proj}.
Nevertheless, the identification of operators that leave $\mathsf{SEP}_{\mathsf{III}}$ is not needed to prove the following Theorem.

\begin{theorem} \label{theoremIII3}
Given two subsets of operators, $\mathcal{A}$ and $\mathcal{B}$, the factorisation condition \eqref{fact}
and Definition \ref{defIII} imply that either $\mathcal{A}$ or $\mathcal{B}$ consists only of operators proportional to the identity.
\end{theorem}
\begin{proof}
Consider separable-III states $|\Psi\rangle=c_X|\Psi_X\rangle+c_Y|\Psi_Y\rangle$, with $|\Psi_X\rangle\in P_X\cdot\mathsf{SEP}_{\mathsf{III}}$, $|\Psi_Y\rangle\in P_Y\cdot\mathsf{SEP}_{\mathsf{III}}$, and $X,Y\in\{LL,LR,RR\}$. The factorisation condition \eqref{fact} reads

\begin{align} \label{fact.IIIXY}
& \displaystyle \sum_{\substack{T,W,W'\in\{X,Y\}\\Z\in\{LL,LR,RR\}}}|c_T|^2\,\overline{c_W}\,c_{W'}\langle\Psi_T|\Psi_T\rangle\langle\Psi_W|A_{W,Z}B_{Z,W'}|\Psi_{W'}\rangle \nonumber \\
& \displaystyle = \sum_{T,T',W,W'\in\{X,Y\}}\overline{c_T}\,c_{T'}\,\overline{c_W}\,c_{W'}\langle\Psi_T|A_{T,T'}|\Psi_{T'}\rangle\langle\Psi_W|B_{W,W'}|\Psi_{W'}\rangle.
\end{align}
Since $c_X$ and $c_Y$ in equation \eqref{fact.IIIXY} are arbitrary, the coefficient of each monomial $\overline{c_T}\,c_{T'}\,\overline{c_W}\,c_{W'}$ must vanish whenever $T\neq T'$ and $W\neq W'$, or $T\neq W'$ and $W\neq T'$. Thus,

\begin{equation} \label{ABXY}
\langle\Psi_X|A_{X,Y}|\Psi_Y\rangle\langle\Psi_X|B_{X,Y}|\Psi_Y\rangle=0.
\end{equation}
Apply now Lemma \ref{lemmaIII} below where the functions therein are $\langle\Psi_X|A_{X,Y}|\Psi_Y\rangle$ and $\langle\Psi_X|B_{X,Y}|\Psi_Y\rangle$, and the variables are the coefficients used to parametrise states $|\Psi_X\rangle$ and $|\Psi_Y\rangle$ (see the discussion after Definition \ref{defIII}). Lemma \ref{lemmaIII} entails that at least one factor in equation \eqref{ABXY} vanish for all separable-III states, i.e. either $\langle\Psi_X|A_{X,Y}|\Psi_Y\rangle=0$ or $\langle\Psi_X|B_{X,Y}|\Psi_Y\rangle=0$ for all $|\Psi_X\rangle\in P_X\cdot\mathsf{SEP}_{\mathsf{III}}$ and $|\Psi_Y\rangle\in P_Y\cdot\mathsf{SEP}_{\mathsf{III}}$. Because $|\Psi_X\rangle$ and $|\Psi_Y\rangle$ span the support of $P_X$ and $P_Y$ respectively, then either $A_{X,Y}=0$ or $B_{X,Y}=0$.

Assume that $A_{X,Y}=0$, as the other case is analogous. Comparing the coefficients of the monomial $|c_T|^2\,\overline{c_X}\,c_Y$ in the right and the left hand sides of equation \eqref{fact.IIIXY}, we obtain

\begin{equation}
\langle\Psi_T|\Psi_T\rangle\langle\Psi_X|A_{X,X}B_{X,Y}|\Psi_Y\rangle \nonumber
=
\langle\Psi_T|A_{T,T}|\Psi_T\rangle\langle\Psi_X|B_{X,Y}|\Psi_Y\rangle,
\end{equation}
for both $T=X$ and $T=Y$. The difference between the two cases ($T=X$ and $T=Y$) of the above equation is

\begin{equation} \label{ABXY3}
\langle\Psi_X|B_{X,Y}|\Psi_Y\rangle\left(\frac{\langle\Psi_X|A_{X,X}|\Psi_X\rangle}{\langle\Psi_X|\Psi_X\rangle}-\frac{\langle\Psi_Y|A_{Y,Y}|\Psi_Y\rangle}{\langle\Psi_Y|\Psi_Y\rangle}\right)=0.
\end{equation}
Lemma \ref{lemmaIII} implies that at least one of the factors in equation \eqref{ABXY3} must be identically zero. If the term in brackets vanishes then $A_{X,X}=\alpha\,P_X$ and $A_{Y,Y}=\alpha\,P_Y$ due to the arbitrariness of the separable-III states $|\Psi_X\rangle$ and $|\Psi_Y\rangle$ (see Lemma \ref{lemmaIII2} below). Therefore, $A=\alpha\mathbbm{1}$ from the arbitrariness of $X,Y\in\{LL,LR,RR\}$, and we have proven Theorem \ref{theoremIII3}. The case $\langle\Psi_X|B_{X,Y}|\Psi_Y\rangle=0$ implies $B_{X,Y}=0$, because separable-III states $|\Psi_X\rangle$ and $|\Psi_Y\rangle$ span the support of $P_X$ and $P_Y$ respectively. At this point, we have proven that $A_{X,Y}=B_{X,Y}=0$ for any $X\neq Y$.

Choose separable-III states $\{|\Psi_Y^j\rangle\}_j$ that form an orthonormal basis of the support of $P_Y$.
Equation \eqref{ABXY} implies

\begin{equation} \label{ABXY2}
0=\sum_j\langle\Psi_X|A_{X,Y}|\Psi_Y^j\rangle\overline{\langle\Psi_X|B_{X,Y}|\Psi_Y^j\rangle}
=\langle\Psi_X|A_{X,Y}B_{Y,X}|\Psi_X\rangle, \qquad \forall\,Y\neq X.
\end{equation}
Using equation \eqref{ABXY2} in the comparison between the coefficients of the monomial $|c_X|^4$ in the left and right hand side of equation \eqref{fact.IIIXY}, we obtain for any $X\in\{LL,LR,RR\}$

\begin{equation}
\langle\Psi_X|\Psi_X\rangle\langle\Psi_X|A_{X,X}B_{X,X}|\Psi_X\rangle
=\langle\Psi_X|A_{X,X}|\Psi_X\rangle\langle\Psi_X|B_{X,X}|\Psi_X\rangle,
\end{equation}

After plugging the above equation in the factorisation \eqref{fact.IIIXY}, and matching the coefficients of $|c_X|^2\,|c_Y|^2$, we get

\begin{equation}
\left(\frac{\langle\Psi_X|A_{X,X}|\Psi_X\rangle}{\langle\Psi_X|\Psi_X\rangle}-\frac{\langle\Psi_Y|A_{Y,Y}|\Psi_Y\rangle}{\langle\Psi_Y|\Psi_Y\rangle}\right)
\left(\frac{\langle\Psi_X|B_{X,X}|\Psi_X\rangle}{\langle\Psi_X|\Psi_X\rangle}-\frac{\langle\Psi_Y|B_{Y,Y}|\Psi_Y\rangle}{\langle\Psi_Y|\Psi_Y\rangle}\right)=0.
\end{equation}
As above, Lemma \ref{lemmaIII} proves that at least one of the factors must identically vanish. According to Lemma \ref{lemmaIII2}, the arbitrariness of separable-III states $|\Psi_X\rangle$ and $|\Psi_Y\rangle$ further implies that either $A_{X,X}=\alpha\,P_X$ or $B_{X,X}=\beta\,P_X$ for any $X\in\{LL,LR,RR\}$. This conclusion, together with the property $A_{X,Y}=B_{X,Y}=0$ for $X\neq Y$ proved above, entails Theorem \ref{theoremIII3}.
\end{proof}

\begin{lemma} \label{lemmaIII}
Given two rational functions $f(x)$ and $g(x)$ of several variables $x=(x_1,x_2,\dots,x_n)$ such that $f(x)g(x)=0$ for all $x$ in the domain of the two functions, then either $f(x)=0$ or $g(x)=0$ for all $x$.
\end{lemma}
\begin{proof}
Fix $x_{j\geqslant2}$ so that the two functions $f(x)=:\tilde f_{x_2,\dots,x_n}(x_1)$ and $g(x)=:\tilde g_{x_2,\dots,x_n}(x_1)$ either are identically zero or have a finite number of zeros $x_1=x_1(x_2,\dots,x_n)$. If both $\tilde f$ and $\tilde g$ have a finite number of zeros, after fixing $x_{j\geqslant2}$, then there are infinitely many n-tuples $x$ such that $f(x)g(x)\neq0$, contradicting the hypothesis of the Lemma.
\end{proof}

\begin{lemma} \label{lemmaIII2}
Given an operator $A$, if for any separable-III states $|\Psi_X\rangle\in P_X\cdot\mathsf{SEP}_{\mathsf{III}}$ and $|\Psi_Y\rangle\in P_Y\cdot\mathsf{SEP}_{\mathsf{III}}$
\begin{equation} \label{diff}
\frac{\langle\Psi_X|A_{X,X}|\Psi_X\rangle}{\langle\Psi_X|\Psi_X\rangle}-\frac{\langle\Psi_Y|A_{Y,Y}|\Psi_Y\rangle}{\langle\Psi_Y|\Psi_Y\rangle}=0,
\end{equation}
then $A_X=\alpha\,P_X$ and $A_Y=\alpha\,P_Y$.
\end{lemma}
\begin{proof}
Equation \eqref{diff} implies that

\begin{align} \label{diff-eq}
\langle\Psi_X|A_{X,X}|\Psi_X\rangle & =\alpha\,\langle\Psi_X|\Psi_X\rangle, \nonumber \\
\langle\Psi_Y|A_{Y,Y}|\Psi_Y\rangle & =\alpha\,\langle\Psi_Y|\Psi_Y\rangle.
\end{align}
with some $\alpha$ independent of the supports of both $P_X$ and $P_Y$. Without loss of generality, we now focus on the first of equations \eqref{diff-eq}, and consider the three cases $X=LL,LR,RR$ with states $|\Psi_X\rangle$ represented as discussed before Theorem \ref{theoremIII3}.

If $X=LL$, separable-III states $|\Psi_X\rangle$ can be represented as in equation \eqref{SI}. Since the state norm is $\langle\Psi_X|\Psi_X\rangle=|c_0|^4+|c_1|^4+|c_0|^2\,|c_1|^2$, and from the arbitrariness of the coefficients $c_{0,1}$, all monomials that appear only on the left-hand-side of equation \eqref{diff-eq}, e.g. $\overline{c_0}^2\,c_1^2$, must be multiplied by a vanishing coefficient. Moreover, the coefficients of the remaining monomials must match between the left and the right hand sides of equation \eqref{diff-eq}. These requirements imply $A_{LL,LL}=\alpha\,P_{LL}$.

The case $X=LR$ is similar, with the difference that separable-III states $|\Psi_{LR}\rangle$ can be represented as $\big(c_0|0\rangle+c_1|1\rangle\big)\otimes\big(d_0|0\rangle+d_1|1\rangle\big)$ and $\langle\Psi_{LR}|\Psi_{LR}\rangle=\big(|c_0|^2+|c_1|^2\big)\big(|d_0|^2+|d_1|^2\big)$. Now, we must compare coefficients of monomials in $c_{0,1}$, $d_{0,1}$, and their complex conjugates. The result is $A_{LR,LR}=\alpha\,P_{LR}$.

The case $X=RR$ is straightforward because all states $|\Psi_{RR}\rangle=P_{RR}\,|\Psi_{RR}\rangle$ are separable-III. Therefore, equation \eqref{diff-eq} must be fulfilled for all bases of the support of $P_{RR}$ that is possible only if $A_{RR,RR}=\alpha\,P_{RR}$.
\end{proof}

Therefore, also entanglement-III is incompatible with any locality notion as sketched in the Introduction: expectations of products of local operators pertaining to different subsystems must factorise for all non-entangled state.
We stress that commutativity between the operator subsets $\mathcal{A}$ and $\mathcal{B}$ have not been used in Theorem \ref{theoremIII3}. This makes Theorem \ref{theoremIII3} a stronger result than analogous theorems for entanglement-I and entanglement-II. We show some examples in the Supplementary Information.

\section{Discussion} \label{disc}

For completeness, we report the other entanglement definitions that correspond to correlations between suitably defined subsystem operators~\cite{review}, generalising thus the Werner's formulation to indistinguishable particles. These definitions are called mode-entanglement and SSR-entanglement.
Mode-entanglement for $N$ bosonic two-level particles is defined by the following.

\begin{definition}[Mode-entanglement~\cite{Zanardi2002-1,Shi2003,Schuch2004-2,Benatti2010}] \label{def-mode}
The set of
pure mode-separable states
is, in first and second quantization respectively,
\begin{align}
\mathsf{SEP}_{\mathsf{mode}} & =\Big\{\mathsf{S}|0\rangle^{\otimes k}\otimes|1\rangle^{\otimes(N-k)}\Big\}_{k=0,1,\dots,N} \\
\widetilde{\mathsf{SEP}}_{\mathsf{mode}} &
=\Big\{\frac{(\mathfrak{a}_0^\dag)^k}{\sqrt{k!}}\,\frac{(\mathfrak{a}_1^\dag)^{N-k}}{\sqrt{(N-k)!}}|\textnormal{vac}\rangle\Big\}_{k=0,1,\dots,N}
\label{SM}
\end{align}
All other pure states are mode-entangled.
\end{definition}
This definition can be generalised to fermions substituting the symmetrization projector $\mathsf{S}$ with the antisymmetrization projector $\mathsf{A}$ or considering anticommuting creation operators: see~\cite{Banuls2007,Benatti2014,review,Szalay2021} for a complete discussion of the fermionic case. If $N=2$, Definition \ref{def-mode} reduces to $\mathsf{SEP}_{\textnormal{mode}}=\big\{|\Phi_0\rangle,|\Phi_1\rangle,|\Phi_2 \rangle\big\}$. Mode-entanglement depends on the choice of the mode basis $\{\mathfrak{a}_0,\mathfrak{a}_1\}$ as it accounts for quantum correlations between modes in the second quantization formalism, and is ubiquitous in quantum optics and quantum field theories, and also applied in several atomic and condensed matter systems (see references in the review~\cite{review}).
The factorisation condition \eqref{fact} is fulfilled if and only if the state $|\Psi\rangle$ therein is mode-separable~\cite{review} when $\mathcal{A}$ consists of all functions of $\mathfrak{a}_0$ and $\mathfrak{a}_0^\dag$ and $\mathcal{B}$ of all functions of $\mathfrak{a}_1$ and $\mathfrak{a}_1^\dag$. Indeed, operators $A\in\mathcal{A}$, $B\in\mathcal{B}$ and $AB$ do not generate mode-entanglement and are the local operators of this theory.

Non-vanishing SSR-entanglement requires that the single particle Hilbert space has at least four dimentions, and so we focus on the same system described for entanglement-III.

\begin{definition}[SSR-entanglement~\cite{Wiseman2003,Ichikawa2010,Sasaki2011}] \label{def-SSR}
The set of
pure SSR-separable states
is, in first and second quantization respectively,
\begin{align}
\mathsf{SEP}_{\mathsf{SSR}}= & \Big\{\sum_{\sigma,\sigma'=0}^1 a_{\sigma,\sigma'}\mathsf{S}|L,\sigma\rangle\otimes|L,\sigma'\rangle+\mathsf{S}\sum_{\sigma=0}^1 b_{\sigma}|L,\sigma\rangle\otimes\sum_{\sigma=0}^1 c_{\sigma}|R,\sigma\rangle
+\sum_{\sigma,\sigma'=0}^1 d_{\sigma,\sigma'}\mathsf{S}|R,\sigma\rangle\otimes|R,\sigma'\rangle\Big\}_{a_{\sigma,\sigma'},b_{\sigma},c_{\sigma},d_{\sigma,\sigma'}\in\mathbbm{C}} \\
\widetilde{\mathsf{SEP}}_{\mathsf{SSR}}= &
\Big\{\sum_{\sigma,\sigma'=0}^1 a_{\sigma,\sigma'}\,\mathfrak{a}_{L,\sigma}^\dag\,\mathfrak{a}_{L,\sigma'}^\dag|\textnormal{vac}\rangle+\sum_{\sigma=0}^1 b_{\sigma}\,\mathfrak{a}_{L,\sigma}^\dag\sum_{\sigma=0}^1 c_{\sigma}\,\mathfrak{a}_{R,\sigma}^\dag|\textnormal{vac}\rangle
+\sum_{\sigma,\sigma'=0}^1 d_{\sigma,\sigma'}\,\mathfrak{a}_{R,\sigma}^\dag\,\mathfrak{a}_{R,\sigma'}^\dag|\textnormal{vac}\rangle\Big\}_{a_{\sigma,\sigma'},b_{\sigma},c_{\sigma},d_{\sigma,\sigma'}\in\mathbbm{C}}
\label{SepSSR}
\end{align}
All other pure states are SSR-entangled.
\end{definition}
SSR-entanglement depends on the choice of the spatial basis $\{|L\rangle,|R\rangle\}$, as entanglement-III does, and, recalling equations \eqref{proj} and \eqref{proj2}, $P_{LR}\mathsf{SEP}_{\mathsf{SSR}}=P_{LR}\mathsf{SEP}_{\mathsf{III}}$. Moreover, SSR-separable states in $P_{LR}\mathsf{SEP}_{\mathsf{III}}$ are isomorphic to $|\sigma\rangle\otimes|\sigma'\rangle$ and indeed represent states of particles effectively distinguished by their spatial localisations.
The name superselection rule entanglement is due to the fact that Definition \ref{def-SSR} is derived within the supports of the projectors $P_{LL}$, $P_{LR}$, $P_{RR}$ (and their many-particle generalisations) and ignoring their superpositions~\cite{Wiseman2003,Ichikawa2010,Sasaki2011,review}. The physical meaning of this superselection rule is that SSR-entanglement recovers standard entanglement when groups of particles can be effectively distinguished. SSR-entanglement is a constrained version of mode-entanglement since $|\Psi\rangle\in\mathsf{SEP}_{\mathsf{SSR}}$ if and only if $P_X|\Psi\rangle\in\mathsf{SEP}_{\mathsf{mode}}$, with $X\in\{LL,LR,RR\}$. Specializing the general argument in Ref.~\cite{review} to the two-particle system described here, the compatibility with locality can be verified when $\mathcal{A}$ is made of all functions of $\mathfrak{a}_{L,\sigma}$ and $\mathfrak{a}_{L,\sigma}^\dag$ and $\mathcal{B}$ of all functions of $\mathfrak{a}_{R,\sigma}$ and $\mathfrak{a}_{R,\sigma}^\dag$, as for mode-entanglement but with the further constraints $P_X\mathcal{A}P_Y=P_X\mathcal{B}P_Y=0$ for all $X\neq Y$, $X,Y\in\{LL,LR,RR\}$. Operators $A\in\mathcal{A}$, $B\in\mathcal{B}$ and $AB$ do not generate SSR-entanglement and are the local operators of this theory.

We analysed different definitions of entanglement for indistinguishable particles in the light of locality.
In particular, entanglement is a form of quantum correlations between subsystems, and therefore entanglement is fully specified only after identifying subsystems. In many of the existing approaches, subsystems are vaguely considered to be particles. We have looked for operators whose correlations correspond to each notion of entanglement. These operators, that define the subsystems by identifying their measurable quantities, form commuting subalgebras, where commutativity entails independence of the subsystems. Nevertheless, we stress that we used neither the commutativity nor the algebra structure in our main proofs. Therefore, our results are more general than what we need for the analysis of entanglement.


The results of our investigation is that three of the five existing entanglement definitions are incompatible with any locality notion formalised as above, because they do not correspond to correlations either between particles or between more general and abstract subsystems. Indeed, for any couple of non-trivial operators there are non-entangled pure states that show correlations.
Therefore, these definitions do not generalise the Werner's formulation of entanglement~\cite{Werner1989}, i.e. the requirement \eqref{fact} for pure separable states, to the domain of indistinguishable particles for any partitioning of the system.
Their practical usefulness may be shown in the framework of different resource theories~\cite{Morris2020} that do not share some of their properties with entanglement theory.
Thus, our results open the way to a deeper investigation of connections between indistinguishable particles entanglement and other resource theories.

Our results are relevant when particle indistinguishability cannot be neglected (see Fig. \ref{fig}). When, on the other hand, particles can be distinguished by means of unambiguos properties, i.e. orthogonal states of certain degrees of freedom like different position eigenstates~\cite{DeMuynck1975,Herbut2001,Herbut2006,Tichy2013,Cunden2014}, then the standard theory of entanglement applies.
Since entanglement of distisguishable particles is a resource for quantum technologies, our analysis shed light into the possibility to identify individually addressable subsystems when particle distinguishability cannot be implemented, as in miniaturised quantum devices with all degrees of freedom employed in the device functioning.

\appendix

\section{Entanglement-I} \label{app:entI}

Definition 1 results from the general definition developed in several approaches~\cite{Paskauskas2001,Eckert2002,Grabowski2011}.
For the sake of completeness, we find conditions for the commutativity of two operators that do not generate entanglement-I. Note that we have not used commutativity in our theorems, although it plays a crucial role the algebraic formulation of entanglement and in the definition of locality \cite{Werner1989,Chitambar2014,review,Sengupta2020}. Then, we prove a variant of Theorem 2 exploiting the commutativity of the operator subsets.
These proofs are derived in first quantisation, but they can be recast in second quantisation, e.g., by mapping the basis (3) into the basis (4).

The matrix in the larger space $\mathbbm{C}^2\otimes\mathbbm{C}^2$ is block diagonal $O\otimes O=\mathsf{S}\big(O\otimes O\big)\mathsf{S}+\mathsf{A}\big(O\otimes O\big)\mathsf{A}$, where $\mathsf{S}$ ($\mathsf{A}$) is the projector onto the (anti-)symmetric subspace. Equation (9) is the matrix representation of the symmetric block, and the antisymmetric block consists of a $1\times 1$ matrix
$a_{00}a_{11}-a_{01}a_{10}$.
Therefore, commutativity between matrices on the symmetric subspace is equivalent to commutativity between the larger matrices $O\otimes O$.

\begin{lemma} \label{theoremI2}
Consider two operators $A=O\otimes O$ and $B=Q\otimes Q$, with Pauli expansions $O=\sum_{\alpha=0}^3x_{\alpha}\sigma_{\alpha}$, and $Q=\sum_{\alpha=0}^3y_{\alpha}\sigma_{\alpha}$, and define $\vec{x}=(x_1,x_2,x_3)$ and $\vec{y}=(y_1,y_2,y_3)$.
Commutativity $[A,B]=0$ implies either $x_0y_0+\vec{x}\cdot\vec{y}=0$ or $\vec{x}\wedge\vec{y}=0$.
\end{lemma}
\begin{proof}
The commutator $[A,B]$ must be identically zero, and so must be each coefficient in the expansion in Pauli matrices, namely $c_{\epsilon,\eta}=\textnormal{Tr}\big([A,B]\sigma_{\epsilon}\otimes\sigma_{\eta}\big)$ ($\epsilon,\eta\in\{0,1,2,3\}$). In particular, if $\epsilon\in\{1,2,3\}$ and $\eta=0$, $c_{\epsilon,\eta}=sz_{\epsilon}$, with $s=x_0y_0+\vec{x}\cdot\vec{y}$ and $z_{1,2,3}$ are components of the external product $\vec{z}=\vec{x}\wedge\vec{y}$. Therefore, either $s=0$ or $\vec{z}=0$ is a necessary condition for $[A,B]=0$.
\end{proof}

\begin{theorem} \label{theoremI3bis}
Given two commuting subsets $\mathcal{A}$ and $\mathcal{B}$ of operators that leave $\mathsf{SEP}_{\mathsf{I}}$ invariant, the factorisation condition (2)
and Definition (1) imply either $\mathcal{A}$ or $\mathcal{B}$ consists only of operators propotional to the identity.
\end{theorem}
\begin{proof}
The commuting subsets $\mathcal{A}$ and $\mathcal{B}$ are formed by operators, $A$ and $B$ respectively, as in Lemma \ref{theoremI2}. Furthermore, separable-I states are $|\Psi\rangle=|\psi\rangle\otimes|\psi\rangle$, and the factorisation condition (2) is equivalent to

\begin{equation} \label{fact.entI2suppl}
\langle\psi|\psi\rangle^2\langle\psi|OQ|\psi\rangle^2=\langle\psi|O|\psi\rangle^2\langle\psi|Q|\psi\rangle^2,
\end{equation}

Let us now consider the two conditions imposed by Lemma \ref{theoremI2}: $s=x_0y_0+\vec{x}\cdot\vec{y}=0$ and $\vec{z}=\vec{x}\wedge\vec{y}=0$.
If $s=0$,
\begin{equation}
OQ=x_0\vec{y}\cdot\vec{\sigma}+y_0\vec{x}\cdot\vec{\sigma}+i\vec{z}\cdot\vec{\sigma}, \quad \vec{\sigma}=(\sigma_1,\sigma_2,\sigma_3).
\end{equation}
If $|\psi\rangle$ is an eigenvector of $\vec{z}\cdot\vec{\sigma}$, then
\begin{equation}
\langle\psi|\psi\rangle^2\langle\psi|OQ|\psi\rangle^2=-\langle\psi|\psi\rangle^4|\vec{z}|^2\leqslant0,
\end{equation}
while, due to the hermiticity of $O$ and $Q$,
\begin{align}
\langle\psi|O|\psi\rangle^2\langle\psi|Q|\psi\rangle^2\geqslant0.
\end{align}
Therefore, the factorisation condition \eqref{fact.entI2suppl} is fulfilled only if $\vec{z}=0$.

Consider now the case $\vec{z}=0$, which implies $\vec{y}=\gamma\vec{x}$. The factorisation condition \eqref{fact.entI2suppl} is a forth order polynomial in $\langle\psi|\vec{x}\cdot\vec{\sigma}|\psi\rangle$. Furthermore, $\langle\psi|\vec{x}\cdot\vec{\sigma}|\psi\rangle$ spans the real axis when $|\psi\rangle$ varies: recall that we have relaxed the normalisation condition for states. Therefore, the coefficient of all powers of $\langle\psi|\vec{x}\cdot\vec{\sigma}|\psi\rangle$ must vanish. In particular, the coefficient of $\langle\psi|\vec{x}\cdot\vec{\sigma}|\psi\rangle^4$ is $\gamma^2$. In conclusion, $\gamma=0$ and thus one of the subset, either $\mathcal{A}$ and $\mathcal{B}$, is made only of operators proportional to the identity.
\end{proof}

\subsection{Examples} \label{exI}

In this section, we provide some examples of the properties proved in the context of entanglement-I. Consider a generic separable-I state $|\Psi\rangle=|\psi\rangle\otimes|\psi\rangle$, with $|\psi\rangle=c_0|0\rangle+c_1|1\rangle$ as in Definition 1, and impose the normalisation $\langle\psi|\psi\rangle=|c_0|^2+|c_1|^2=1$. First of all, the action of single-particle operators $O\otimes\mathbbm{1}$ do not preserve the particle permutation symmetry. Indeed, the resulting state is not invariant under the exchange of particles:

\begin{equation}
\big(O\otimes\mathbbm{1}\big)|\psi\rangle\otimes|\psi\rangle=\big(O|\psi\rangle\big)\otimes|\psi\rangle.
\end{equation}
On the other hand, symmetrised single-particle operators $O\otimes\mathbbm{1}+\mathbbm{1}\otimes O$ generate entanglement-I:

\begin{equation} \label{sym1part-op}
\big(O\otimes\mathbbm{1}+\mathbbm{1}\otimes O\big)\psi\rangle\otimes|\psi\rangle=\big(O|\psi\rangle\big)\otimes|\psi\rangle+|\psi\rangle\otimes\big(O|\psi\rangle\big).
\end{equation}
The state \eqref{sym1part-op} is not separable-I. If, e.g., either $c_0=1$ or $c_1=1$ and $O=\sigma_1$ is the first Pauli matrix, the state \eqref{sym1part-op} is $|0\rangle\otimes|1\rangle+|1\rangle\otimes|0\rangle$.

As examples of commuting operators that do not generate entanglement-I, consider $O\otimes O$ and $Q\otimes Q$, with Pauli matrices $O=\sigma_1$ and $Q=\sigma_2$. Therefore,

\begin{equation}
\langle\Psi|\big(O\otimes O\big)\big(Q\otimes Q\big)|\Psi\rangle=-\langle\psi|\sigma_3|\psi\rangle^2=-\big(|c_0|^2-|c_1|^2\big)^2,
\end{equation}
while

\begin{align}
\langle\Psi|\big(O\otimes O\big)|\Psi\rangle & =\langle\psi|\sigma_1|\psi\rangle^2=4\big(\textnormal{Re}(\overline{c_0}c_1)\big)^2, \\ 
\langle\Psi|\big(Q\otimes Q\big)|\Psi\rangle & =\langle\psi|\sigma_2|\psi\rangle^2=4\big(\textnormal{Im}(\overline{c_0}c_1)\big)^2.
\end{align}
The factorisation condition is violated,

\begin{equation} \label{ineqI}
\langle\Psi|\big(O\otimes O\big)\big(Q\otimes Q\big)|\Psi\rangle
\neq\langle\Psi|\big(O\otimes O\big)|\Psi\rangle\langle\Psi|\big(Q\otimes Q\big)|\Psi\rangle.
\end{equation}
For instance, if either $c_0=1$ or $c_1=1$ the inequality \eqref{ineqI} reads $-1\neq0$, and if $c_0=1/\sqrt{5}, c_1=2/\sqrt{5}$ we obtain $-\frac{9}{25}\neq0$.

\section{Entanglement-II} \label{app:entII}

Definition 2 follows from the general frameworks in Refs.~\cite{Herbut2001,Schliemann2001,Paskauskas2001,Eckert2002,
Plastino2009,
Grabowski2011,Li2001,
Ghirardi2002,
Iemini2013}. It is worthwhile to stress that separable-I states are also separable-II, but there are separable-II states that are entangled-I.
Theorem 4 shows the incompatibility of entanglement-II with the locality notion, exploiting similar results for entanglement-I. Now, we provide a more detailed argument that relies on the explicit forms of operators that leave $\mathsf{SEP}_{\mathsf{II}}$ invariant.

\begin{theorem} \label{oplocII}
Any operator that leaves $\mathsf{SEP}_{\mathsf{II}}$ invariant either leaves $\mathsf{SEP}_{\mathsf{II}}\setminus\mathsf{SEP}_{\mathsf{I}}$ invariant or sends $\mathsf{SEP}_{\mathsf{II}}$ in $\mathsf{SEP}_{\mathsf{I}}$.
\end{theorem}

\begin{proof}
Consider an operator $A$ that leaves $\mathsf{SEP}_{\mathsf{II}}$ invariant, and a state $|\Psi\rangle\in\mathsf{SEP}_{\mathsf{II}}\setminus\mathsf{SEP}_{\mathsf{I}}$. Either $A|\Psi\rangle\in\mathsf{SEP}_{\mathsf{I}}$ or $A|\Psi\rangle\in\mathsf{SEP}_{\mathsf{II}}\setminus\mathsf{SEP}_{\mathsf{I}}$. A necessary and sufficient condition for $A|\Psi\rangle\in\mathsf{SEP}_{\mathsf{I}}$ is equation (8), as in the proof of Theorem 3,

\begin{equation} \label{PA2}
P_A(c_0,\overline{c_0},c_1,\overline{c_1}):=\langle\Phi_2|A|\Psi\rangle^2-2\,\langle\Phi_0|A|\Psi\rangle\langle\Phi_1|A|\Psi\rangle=0.
\end{equation}
The crucial difference with the case $|\Psi\rangle\in\mathsf{SEP}_{\mathsf{I}}$ of Theorem 3 is that the polynomial $P_A$ now depends also on complex conjugates $\overline{c_0}$ and $\overline{c_1}$.
A necessary and sufficient condition for $A|\Psi\rangle\in\mathsf{SEP}_{\mathsf{II}}\setminus\mathsf{SEP}_{\mathsf{I}}$ is equation (22), exactly as in the proof of Theorem 3.

Now, also equation \eqref{PA2} can have no solutions. Therefore, fixing $c_0$ and $\textnormal{Im}\,c_1$, each of equation \eqref{PA2} and equation (22) either have a finite number (possibly zero) of solutions or are tautologies. As in the proof of Theorem 3, if both equations have finitely many solutions, there are states in $\mathsf{SEP}_{\mathsf{II}}$ that are send out of $\mathsf{SEP}_{\mathsf{II}}$, contradicting the hypothesis. Therefore, one equation is a tautology. For the other equation there are two possibilities: it has no solutions otherwise there are states sent to both $\mathsf{SEP}_{\mathsf{I}}$ and $\mathsf{SEP}_{\mathsf{II}}\setminus\mathsf{SEP}_{\mathsf{I}}$ contradicting the linearity of $A$, or $A$ annihilates any solution $|\Psi\rangle$.

If equation \eqref{PA2} is the tautology, then $A\cdot\big(\mathsf{SEP}_{\mathsf{II}}\setminus\mathsf{SEP}_{\mathsf{I}}\big)\subset\mathsf{SEP}_{\mathsf{I}}$ which implies, together with Theorem 3, $A\cdot\mathsf{SEP}_{\mathsf{II}}\subset\mathsf{SEP}_{\mathsf{I}}$. If equation (22) is the tautology, then $A\cdot\big(\mathsf{SEP}_{\mathsf{II}}\setminus\mathsf{SEP}_{\mathsf{I}}\big)\subset\mathsf{SEP}_{\mathsf{II}}\setminus\mathsf{SEP}_{\mathsf{I}}$.
\end{proof}

\begin{theorem} \label{locopII.repr}
The operators that leave $\mathsf{SEP}_{\mathsf{II}}$ invariant are represented on $\mathbbm{C}^2\otimes\mathbbm{C}^2$ as $O\otimes O$ where
$O$ is proportional to a unitary matrix.
\end{theorem}
\begin{proof}
From Theorem 3, the operators considered here leave $\mathsf{SEP}_{\mathsf{I}}$ invariant, and are therefore represented as $O\otimes O$ on the larger Hilbert space $\mathbbm{C}^2\otimes\mathbbm{C}^2$, by Theorem 1. Consider a separable-II state

\begin{equation}
|\Psi\rangle=\mathsf{S}\,|\psi\rangle\otimes|\psi^\perp\rangle\in\mathsf{SEP}_{\mathsf{II}}\setminus\mathsf{SEP}_{\mathsf{I}},
\end{equation}
with $\langle\psi|\psi^\perp\rangle=0$.

If $O\otimes O$ sends $|\Psi\rangle$ to $\mathsf{SEP}_{\mathsf{I}}$, namely

\begin{equation}
O\otimes O|\Psi\rangle=\mathsf{S}\big(O|\psi\rangle\otimes O|\psi^\perp\rangle\big)\in\mathsf{SEP}_{\mathsf{I}},
\end{equation}
then $O|\psi\rangle=O|\psi^\perp\rangle$ which implies $O=\lambda|+\rangle\langle+|$ with $|+\rangle=|\psi\rangle+|\psi^\perp\rangle$. This result must hold for any basis $\big\{|\psi\rangle,|\psi^\perp\rangle\big\}$, according to Theorem \ref{oplocII}. Therefore, $\lambda$ must be zero, and there are no operators that transform $\mathsf{SEP}_{\mathsf{II}}\setminus\mathsf{SEP}_{\mathsf{I}}$ to $\mathsf{SEP}_{\mathsf{I}}$.

The other possibility left by Theorem \ref{oplocII} is that

\begin{equation}
O\otimes O|\Psi\rangle=\mathsf{S}\big(O|\psi\rangle\otimes O|\psi^\perp\rangle\big)\in\mathsf{SEP}_{\mathsf{II}}\setminus\mathsf{SEP}_{\mathsf{I}}
\end{equation}
for any basis $\big\{|\psi\rangle,|\psi^\perp\rangle\big\}$. From the Definition 2,

\begin{equation}
\langle\psi|O^\dag O|\psi^\perp\rangle=0, \qquad \forall\,\big\{|\psi\rangle,|\psi^\perp\rangle\big\}.
\end{equation}
The latter equation implies that $O^\dag O$ is diagonal in any basis, thus it must be proportional to the identity matrix. Therefore, $O$ is proportional to a unitary matrix.
\end{proof}

\begin{corollary}
The operators that leave $\mathsf{SEP}_{\mathsf{II}}$ invariant do not form an algebra.
\end{corollary}
\begin{proof}
If $O\otimes O$ is in the algebra, also $O^\dag O\otimes O^\dag O\propto\mathbbm{1}$ and thus the identity are in the algebra. Nevertheless, linear combinations of $O\otimes O$ and the identity are not of the form required by Theorem \ref{locopII.repr}.
\end{proof}

In conclusion, it is not possible to identity an algebra of operators that do not generate entanglement-II, and this notion of entanglement is not compatible with the existence of local subsystems defined by subalgebras of their local operators.

\subsection{Examples} \label{exII}

In this section, we discuss some examples of the above theorems for entanglement-II.
We remind that separable-I states are also separable-II (see Definition 2), and that operators that do not generate entanglement-II are special cases of those that do not gneerate entanglement-I (see Theorem \ref{locopII.repr}). In particular, the examples discussed for entanglement-I applies also to the framework of entanglement-II. The only difference is that it is not enough to compare states \eqref{sym1part-op}, that result from the action of symmetrised single-particle operators $O\otimes\mathbbm{1}+\mathbbm{1}\otimes O$, with separable-I states, but rather with the larger class of separable-II states. Nevertheless, the condition that the state \eqref{sym1part-op} is separable-II, although not separable-I, reads $\langle\psi|O|\psi\rangle=0$ (see Definition 2). This condition, together with the arbitrariness of the state $|\psi\rangle$, implies that $O$ is the zero operator. Therefore, symmetrised single-particle operators $O\otimes\mathbbm{1}+\mathbbm{1}\otimes O$ generate entanglement-II.

We now complement the examples in the previous section with others that exploit separable-II states that are not separable-I, namely

\begin{equation}
|\Psi\rangle=\frac{1}{\sqrt{2}}\big(|\psi\rangle\otimes|\psi^\perp\rangle+|\psi^\perp\rangle\otimes|\psi\rangle\big),
\end{equation}
where

\begin{align}
|\psi\rangle & =c_0|0\rangle+c_1|1\rangle, \\
|\psi^\perp\rangle & =\overline{c_1}|0\rangle-\overline{c_0}|1\rangle,
\end{align}
with the normalisation condition $|c_0|^2+|c_1|^2=1$.
Consider, as in section \ref{exI}, the operators $O\otimes O$ and $Q\otimes Q$ with $O=\sigma_1$ and $Q=\sigma_2$, that do not generate entanglement-II. The expectation values of these operators are

\begin{equation}
\langle\Psi|\big(O\otimes O\big)\big(Q\otimes Q\big)|\Psi\rangle=
-\langle\psi|\sigma_3|\psi\rangle\langle\psi^\perp|\sigma_3|\psi^\perp\rangle-\big|\langle\psi^\perp|\sigma_3|\psi\rangle\big|^2=
|c_0|^4+|c_1|^4-6\,|c_0\,c_1|^2,
\end{equation}
and

\begin{align}
\langle\Psi|\big(O\otimes O\big)|\Psi\rangle= & \big(\langle\psi|\sigma_1|\psi\rangle\langle\psi^\perp|\sigma_1|\psi^\perp\rangle+\big|\langle\psi^\perp|\sigma_1|\psi\rangle\big|^2\big)
=\Big(\big|c_1^2-c_0^2\big|^2-4\big(\textnormal{Re}(\overline{c_0}c_1)\big)^2\Big), \\
\langle\Psi|\big(Q\otimes Q\big)|\Psi\rangle= & \big(\langle\psi|\sigma_2|\psi\rangle\langle\psi^\perp|\sigma_2|\psi^\perp\rangle+\big|\langle\psi^\perp|\sigma_2|\psi\rangle\big|^2\big)
=\Big(\big|c_0^2+c_1^2\big|^2-4\big(\textnormal{Im}(\overline{c_0}c_1)\big)^2\Big).
\end{align}
Therefore, the factorisation condition (2) is violated in general, as in equation \eqref{ineqI}.

\section{Entanglement-III} \label{app:entIII}

Note that Definition 3 is called entanglement-IV in~\cite{review} whereas entanglement-III therein is the definition called SSR-entanglement.

In order to define separable-III states in Definition 3, we need to introduce some preliminary notions. First of all, define reductions from the two-particle Hilbert space to the single-particle Hilbert space.

\begin{equation} \label{Pi-psi}
\Pi_{\psi} \, \mathsf{S}\,|\phi\rangle\otimes|\zeta\rangle=\langle\psi|\phi\rangle|\zeta\rangle+\eta\langle\psi|\zeta\rangle|\phi\rangle\ ,
\end{equation}
where $\eta=+1(-1)$ for bosons (fermions). Within this approach, a so-called reduced single-particle density matrix is defined relative to a single-particle subspace $\mathcal{K}$:

\begin{equation} \label{rho1}
\rho^{(1)}=\frac{1}{\displaystyle \sum_{\substack{k\,:\,\{|\psi_k\rangle\}_k\\\textnormal{ONB of }\mathcal{K}}}\big|\big|\Pi_{\psi_k}|\Psi\rangle\big|\big|^2}\sum_{\substack{k\,:\,\{|\psi_k\rangle\}_k\\\textnormal{ONB of }\mathcal{K}}}\Pi_{\psi_k}|\Psi\rangle\langle\Psi|\Pi_{\psi_k}^\dag,
\end{equation}
where $\big\{|\psi_k\rangle\big\}_k$ is any orthonormal basis (ONB) of $\mathcal{K}$.
In second quantisation, $\rho^{(1)}$ reads
\begin{equation} \label{rho1.2}
\rho^{(1)}=\frac{1}{\displaystyle \sum_{\substack{k\,:\,\{|\psi_k\rangle\}_k\\\textnormal{ONB of }\mathcal{K}}}\big|\big|\mathfrak{a}_{\psi_k}|\Psi\rangle\big|\big|^2}\sum_{\substack{k\,:\,\{|\psi_k\rangle\}_k\\\textnormal{ONB of }\mathcal{K}}}\mathfrak{a}_{\psi_k}|\Psi\rangle\langle\Psi|\mathfrak{a}_{\psi_k}^\dag,
\end{equation}
where $\mathfrak{a}_{\psi_k}^\dag$ ($\mathfrak{a}_{\psi_k}$) creates (annihilates) a particle in the state $|\psi_k\rangle$, and $\big[\mathfrak{a}_{\psi_k},\mathfrak{a}_{\psi_{k'}}^\dag\big]=\delta_{k,k'}$.
Note that the density matrix $\rho^{(1)}$ does not reproduce expectations of single-particle observables \cite{Benatti2017}, and, yet, is used to define entanglement.

A two-particle state $|\Psi\rangle$ is said to be separable-III if $\rho^{(1)}$ is a one-dimensional projector, i.e. if $\big(\rho^{(1)}\big)^2=\rho^{(1)}$. Since $\mathcal{K}=\textnormal{span}\big\{|L,\sigma\rangle\big\}_{\sigma}$, the states $|\psi_k\rangle$ in equation \eqref{rho1} are of the form $|L,\sigma\rangle$ for some internal state $\sigma$. The condition $\big(\rho^{(1)}\big)^2=\rho^{(1)}$ implies that $\rho^{(1)}$ has only one non zero eigenvalue, and so there exists a basis $\big\{|L,\sigma\rangle,|L,\sigma^\perp\rangle\big\}$, with $\langle\sigma|\sigma^\perp\rangle=0$, such that $\Pi_{L,\sigma^\perp}|\Psi\rangle=0$. Therefore, separable-III states can be expressed with single-particle states with components along the states $|L,\sigma\rangle$, $|R,\sigma\rangle$, $|R,\sigma^\perp\rangle$, but not along $|L,\sigma^\perp\rangle$. Definition 3 follows from the above considerations.

\subsection{Examples}

In this section, we present some examples for the impossibility to reconcile the definition of entanglement-III with the Werner's formulation. First of all, consider separable-III states with two particles in the left location, namely separable-III states that lie in the support of the projector $P_{LL}$ defined in equations (25). The support of $P_{LL}$ is spanned by states where only internal degrees of freedom varies. Therefore, this subspace is isomorphic to $\mathbbm{C}^2\otimes\mathbbm{C}^2$, and entanglement-III in this subspace is the same as entanglement-I. Therefore, all the examples shown in section \ref{exI} applies also here.

Consider now more general examples of separable-III states:

\begin{equation}
|\Psi\rangle=c_{LL}|L,0\rangle^{\otimes 2}+c_{LR}\,\mathsf{S}\,|L,0\rangle\otimes|R,1\rangle+c_{RR}|R,1\rangle^{\otimes 2},
\end{equation}
in first quantization, or

\begin{equation}
|\Psi\rangle=c_{LL}\,\frac{\big(\mathfrak{a}_{L,0}^\dag\big)^2}{\sqrt{2}}|\textnormal{vac}\rangle+c_{LR}\,\mathfrak{a}_{L,0}^\dag \mathfrak{a}_{R,1}^\dag|\textnormal{vac}\rangle
+c_{RR}\,\frac{\big(\mathfrak{a}_{R,1}^\dag\big)^2}{\sqrt{2}}|\textnormal{vac}\rangle,
\end{equation}
in second quantization, with the normalisation $\langle\Psi|\Psi\rangle=|c_{LL}|^2+|c_{LR}|^2+|c_{RR}|^2=1$.
Note that thess state are separable-III for both choices of the single-particle subspace $\mathcal{K}=\textnormal{span}\{|L,0\rangle,|L,1\rangle\}$ and $\mathcal{K}=\textnormal{span}\{|R,0\rangle,|R,1\rangle\}$ (see the comments before Definition 3 and equations \eqref{rho1} and \eqref{rho1.2}). Consider also the operators

\begin{align}
A & =|L\rangle\langle L|\otimes\sigma_3\otimes\mathbbm{1}+\mathbbm{1}\otimes|L\rangle\langle L|\otimes\sigma_3, \\
B & =|R\rangle\langle R|\otimes\sigma_3\otimes\mathbbm{1}+\mathbbm{1}\otimes|R\rangle\langle R|\otimes\sigma_3,
\end{align}
where $\mathbbm{1}$ is the identity matrix on the single-particle Hilbert space with spatial and internal degrees of freedom. These operators can be written in second quantization as

\begin{align}
A & =a_{L,0}^\dag a_{L,0}-a_{L,1}^\dag a_{L,1}, \\
B & =a_{R,0}^\dag a_{R,0}-a_{R,1}^\dag a_{R,1},
\end{align}
and are considered local within the theory of entanglement-III \cite{LoFranco2016,Castellini2018}. Expectation values are

\begin{align}
& \langle\Psi|AB|\Psi\rangle=-|c_{LR}|^2, \\
& \langle\Psi|A|\Psi\rangle=2\,|c_{LL}|^2+|c_{LR}|^2, \\
& \langle\Psi|B|\Psi\rangle=-2\,|c_{RR}|^2-|c_{LR}|^2.
\end{align}
Therefore, the factorisation condition is violated:

\begin{equation} \label{ineqIII}
\langle\Psi|AB|\Psi\rangle\neq\langle\Psi|A|\Psi\rangle\langle\Psi|B|\Psi\rangle.
\end{equation}
For instance, if $c_{LR}=0$ inequality \eqref{ineqIII} reads $0\neq-4|c_{LL}|^2|c_{RR}|^2$, and if $c_{RR}=0$ we obtain $1\neq1+|c_{LL}|^2$. The factorisation condition is fulfilled only if $c_{LL}=c_{RR}=0$, and thus $c_{LR}=1$ which implies that the states $|\Psi\rangle$ lie in the support of the projector $P_{LR}$. Within this subspace, entanglement-III is equivalent to SSR-entanglement, studied in \cite{Wiseman2003,Ichikawa2010,Sasaki2011} (see also the section \ref{disc}), that is compatible with the Werner's formulation and is a special case of mode-entanglement \cite{review}.

\section*{Acknowledgments}

U.~M. is financially supported by the European Union's Horizon 2020 research and innovation programme under the Marie Sk\l odowska-Curie grant agreement No. 754496 - FELLINI. T.~J.~F.~J. and U.~M. acknowledges the DAAD RISE Worldwide programme reference code HR-PH-4248.


%

\end{document}